\documentclass[aip,apl,groupedaddress,reprint]{revtex4-1}
\usepackage{graphicx}   
\usepackage{siunitx}
\usepackage[version=4]{mhchem}
\usepackage[hidelinks]{hyperref}   
\usepackage[capitalize, nameinlink]{cleveref}

\begin{document}
\title{Towards high all-optical data writing rates in synthetic ferrimagnets}
\date{\today}

\author{Youri L.W. van Hees}
\email{y.l.w.v.hees@tue.nl}
\affiliation{Department of Applied Physics, Institute for Photonic Integration, Eindhoven University of Technology, P.O. Box 513 5600 MB Eindhoven, the Netherlands}

\author{Bert Koopmans}
\affiliation{Department of Applied Physics, Institute for Photonic Integration, Eindhoven University of Technology, P.O. Box 513 5600 MB Eindhoven, the Netherlands}

\author{Reinoud Lavrijsen}
\affiliation{Department of Applied Physics, Institute for Photonic Integration, Eindhoven University of Technology, P.O. Box 513 5600 MB Eindhoven, the Netherlands}

\begin{abstract}
Although all-optical magnetization switching with fs laser pulses has garnered much technological interest, the ultimate data rates achievable have scarcely been investigated.
Recently it has been shown that after a switching event in a GdCo alloy, a second laser pulse arriving 7 ps later can consistently switch the magnetization.
However, it is as of yet unknown whether the same holds in layered ferrimagnetic systems, which hold much promise for applications.
In this work we investigate the minimum time delay required between two subsequent switching events in synthetic ferrimagnetic Co/Gd bilayers using two fs laser pulses.
We experimentally demonstrate that the minimum time delay needed for consistent switching can be as low as 10 ps.
Moreover, we demonstrate the importance of engineering heat diffusion away from the magnetic material, as well as control over the laser pulse power.
This behavior is reproduced using modelling, where we find that the second switch can occur even when the magnetization is not fully recovered.
We further confirm that heat diffusion is a critical factor in reducing the time delay for the second switch, while also confirming a critical dependence on laser power.
\end{abstract}

\maketitle
\begin{figure}[t]
    \includegraphics[width=234pt]{{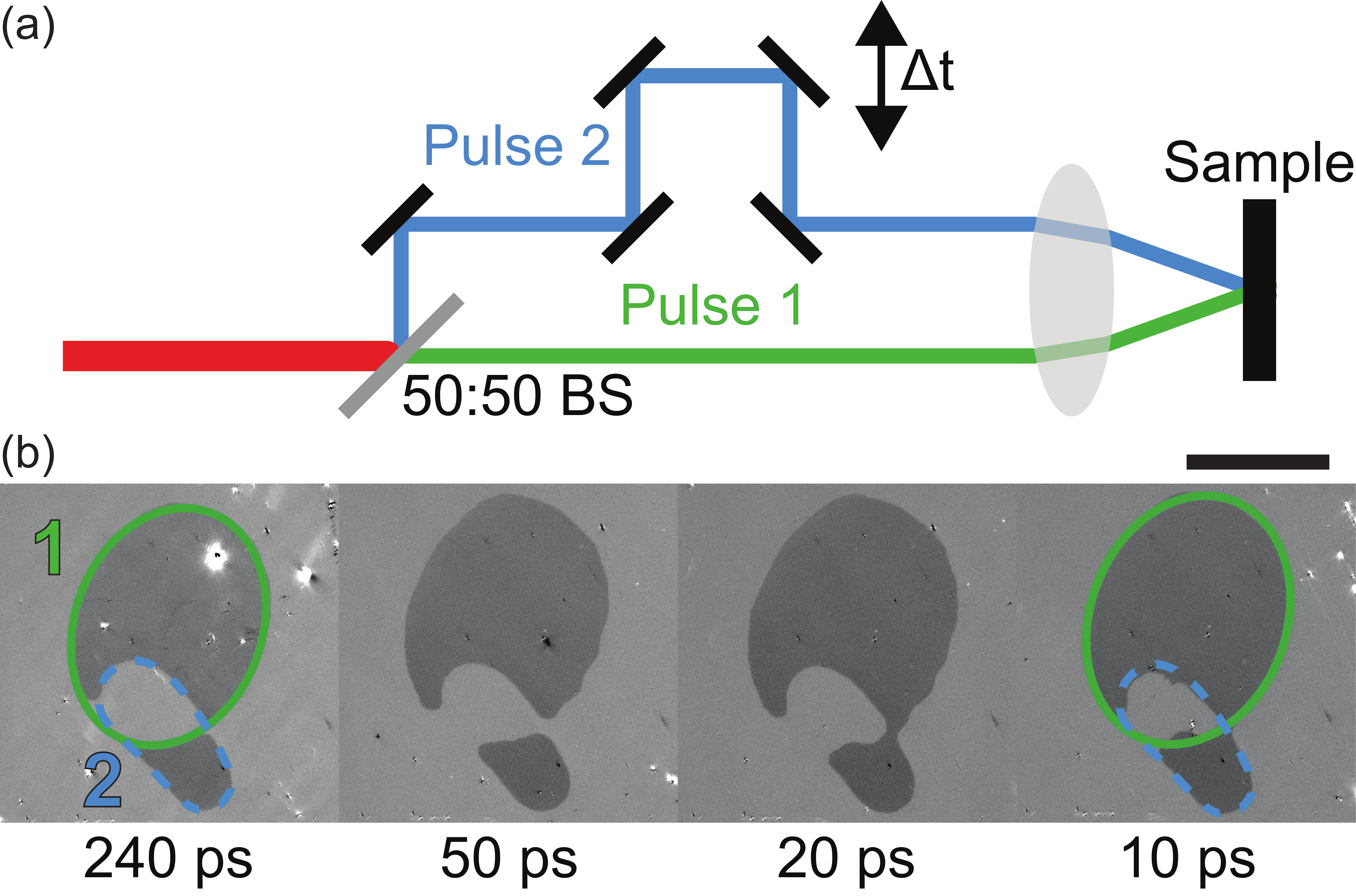}}
    \caption{(a) Sketch of the experimental setup, colors are for illustrative purposes only. (b) Kerr microscopy images of the magnetic state of a Si:B//Ta(4)/Pt(4)/Co(1)/Gd(3)/Ta(4) sample after exposure to two $\sim$ 100 fs laser pulses with varying time delay. The green and blue lines enclose the areas that would be switched by the two pulses individually. The scale bar represents 50 {\textmu}m.}
    \label{fig:fig1}
\end{figure}

All-optical switching (AOS) of the magnetization of thin film ferrimagnets using single femtosecond laser pulses has been demonstrated to be a robust, ultrafast, and energy efficient method to write data with promise for future memory devices\cite{kimel2019review}.
The mechanism was first discovered in ferrimagnetic GdFeCo alloys\cite{Stanciu2007GdFeCo,Ostler2012GdFeCo,leguyader2012nanostructures,khorsand2012MCD,gorchon2016TeTp}, and was soon followed by demonstrations in synthetic ferrimagnets (Co/Gd and Co/Tb)\cite{Lalieu2017CoGd,Aviles2019CoTb}.
The switching was shown to be symmetrical, with each subsequent laser pulse toggling the magnetization between the 'up' and 'down' states, a process which can be repeated over hundreds of millions of cycles without failure\cite{peeters2022influence}.
The energy efficiency of AOS is especially interesting for applications, with typical energies of only tens of fJ needed to switch nanoscale bits\cite{savoini2012highly,Lalieu2017CoGd}.
Moreover, the magnetization only takes a few picoseconds to cross zero\cite{Radu2011GdFeCo,ceballos2021role}, potentially implying near THz writing speeds.
However, this is not the most relevant timescale for determining the ultimate data rates, as it is expected that the magnetization should relax to the opposite state after each switching event to facilitate the next switch.
This remagnetization process can potentially take hundreds of ps\cite{peeters2022influence}, and is expected to be governed by heat diffusion away from the magnetic layers\cite{Lisowski2005GdDynamics}.
Despite the large body of research into AOS, understanding of the ultimate speed with which subsequent switches can actually take place remains scarce\cite{atxitia2018ultrafast}.
It has been shown experimentally that a second fs laser pulse can consistently switch the magnetization in a GdFeCo alloy again after 300 to 400 ps\cite{wang2021dual}.
Here it was conjectured that this is likely not a fundamental limit of the switching process but rather a limit imposed by heat diffusion.
Very recently it was demonstrated by Steinbach et al. that this is indeed the case, with a minimum possible waiting time of 7 ps in a GdCo alloy when a substrate with proper heat conductivity is used\cite{steinbach2022accelerating}.
The remagnetization time of the alloy was emphasized by the authors as a critical factor for facilitating the second switch.
In this light, one might expect longer waiting times between switches in technologically highly relevant synthetic ferrimagnets\cite{pham2016dwmotion,blasing2018exchange,wang2020annealing}, where the magnetic sublattices are less strongly coupled and therefore expected to remagnetize slower after AOS than in alloys\cite{beens2019intermixing}.
\\ \\
In this work we experimentally demonstrate that by taking into account the heat diffusion of the sample, switching events can take place with very short delays even in synthetic ferrimagnetic Co/Gd systems.
The minimum time delay for the second pulse to consistently switch the magnetization is found to be as low as 10 ps, yielding potential writing rates of up to 100 GHz.
The importance of the heat diffusion in this process is highlighted by demonstrating a larger minimum time for substrates with lower heat conductivity.
Moreover, the absence of rapid double switching when reversing the order of the laser pulses or when slightly increasing the power of one of the pulses highlights the need for careful control of the irradiation conditions.
Finally, we present modelling results using the Microscopic Three Temperature Model (M3TM)\cite{beens2019comparing} confirming that heat diffusion is the dominant factor in reducing the delay for the second switch, and also illuminating the critical role of the laser pulse energy.
\\ \\
The experiments in this work are performed on Ta(4)/Pt(4)/Co(1)/Gd(3)/Cap.(4) multilayer stacks (where numbers in parentheses indicate layer thickness in nm, and `Cap.' is Ta or TaN) known to exhibit AOS\cite{Lalieu2017CoGd}, which are deposited on Si substrates using DC magnetron sputter deposition.
In order to achieve a high heat conductivity, a degenerately Boron doped substrate is used.
As sketched in Fig. \ref{fig:fig1}a, individual $\sim$ 100 fs laser pulses with a central wavelength of 700 nm are split in two using a 50:50 beam splitter, after which one pulse goes through a delay line so that the time delay between the two pulses can be adjusted.
Both pulses are subsequently focused onto the sample via the same objective.
The magnetic state of a sample after exposure is imaged using an ex situ Kerr microscope.
The magnetic state of the sample after exposing different regions to sets of two pulses with varying time delays is shown in Fig. \ref{fig:fig1}b.
Here, the regions that would be switched by the first and second pulse separately are indicated by the solid green and dashed blue shapes, respectively.
For the longest time delay shown here (240 ps) a clear region within the overlap of the two pulses is observed where the magnetization is switched twice, returning to the initial state.
This is comparable to previous work on GdFeCo alloys, where the second switch was possible after 300-400 ps\cite{wang2021dual}.
Moreover, comparable to recent work on GdCo\cite{steinbach2022accelerating} we find that double switching also occurs when reducing the time delay to 50 ps (as indicated by the red dotted circle), and even stays possible for time delays as low as 10 ps.
As will be discussed later, this result is somewhat surprising, as magnetization recovery is expected to be faster in alloys than in synthetic ferrimagnets, where the magnetic sublattices are only coupled at the interface, and the Gd magnetization shows a stronger temperature dependence.
Comparing the images for 240 and 10 ps time delay, we note the shrinkage of the region where double switching occurs for the shortest time delay, indicating a more critical dependence on the exact laser fluence of the two pulses.
Switching in the region where the second pulse is below threshold, as was observed in similar experiments on GdFeCo\cite{wang2021dual}, has not been observed in the present work.
\\ \\
To investigate the effect of heat diffusion on the minimum double switching time, we now turn to a substrate with lower heat conductivity.
A Ta(4)/Pt(4)/Co(1)/Gd(3)/TaN(4) stack is deposited on a silicon substrate with a 100 nm coating of SiO\textsubscript{2}.
The oxide layer is expected to be less efficient in conducting heat away from the metallic multilayer stack than the semimetallic Si:B substrate.
Remagnetization therefore is expected to be slower, and the minimum time needed between two pulses for consistent switching should be larger.
\begin{figure}[t]
    \includegraphics[width=234pt]{{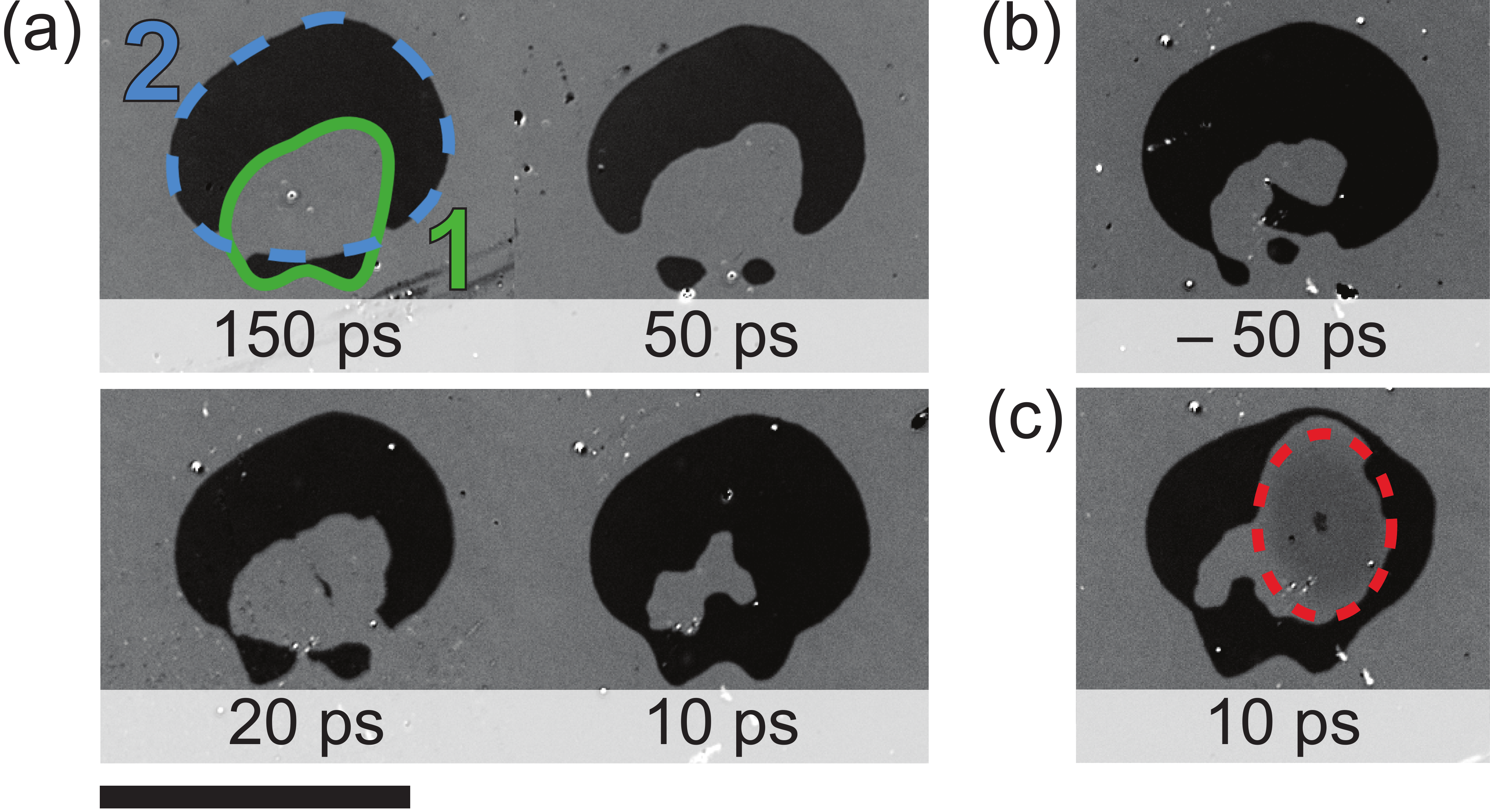}}
    \caption{Kerr microscopy images of the magnetic state of a Si/SiO\textsubscript{2}/Ta(4)/Pt(4)/Co(1)/Gd(3)/TaN(4) sample after exposure to two $\sim$ 100 fs laser pulses with varying time delay. The solid green and dashed blue lines enclose the areas that would be switched by the two pulses individually. The scale bar represents 50 {\textmu}m. In (a) and (c), pulse 1 arrives before pulse 2, whereas in (b) pulse 2 arrives first. The intensity of pulse 2 is 10\% higher in (c) than in (a) and (b). The red dotted ellipse in (c) indicates the area where the effect of the combined heating of both pulses was high enough to induce magnetic and/or structural damage.}
    \label{fig:fig2}
\end{figure}
We perform the same experiment presented in Fig. \ref{fig:fig1} on the multilayer stack grown on the Si/SiO\textsubscript{2} substrate, the result of which is shown in Fig. \ref{fig:fig2}a.
The areas switched by the two pulses separately are again indicated with solid green and dashed blue lines in the Kerr microscope image for 150 ps time delay.
Note that the intensity of the two laser pulses is identical, however the power density in the first pulse is higher, leading to a smaller switched area.
Consistent double switching is again observed for time delays down to 20 ps.
However for a time delay of 10 ps, where double switching was possible previously, different behavior is observed.
Only a small part of the area where both pulses overlapped has returned to the initial state, with this area being rather complexly bounded.
This is an indication that this area is actually not consistently switched twice, but rather remagnetized in a random state after cooling down.
Such a process can occur when the temperature of the lattice exceeds the Curie temperature for a longer time, leading to a complete loss of magnetization for a short period of time\cite{gorchon2016TeTp}.
This observation is consistent with the expectation that heat remains in the system for a longer time when heat transfer is impeded by the oxide coating.
\\ \\
To illustrate the sensitivity of the double switching process, Fig. \ref{fig:fig2}b shows the result of exposure when the order of the two pulses is swapped.
For a delay as large as 50 ps, a more random state is found in the overlapping region, indicating a critical dependence on the power distribution of the two laser pulses.
More specifically, when the pulse with higher intensity ('pulse 1') arrives last, double switching is less consistent.
This can again be understood by realizing that by strongly heating the sample with the second pulse before the heat from the first pulse has significantly dissipated, the lattice temperature can exceed the Curie temperature.
As an additional demonstration of the criticality of the laser power, we increase the intensity of pulse 1 by as little as 10\%, leading to the magnetic state shown in \ref{fig:fig2}c.
Here we find a region (indicated by the red dashed ellipse) where the magnetic properties of the sample have changed due to laser irradiation.
Both the small dark dot in the center of this region as well as the lighter area around it have been annealed by the combined effect of both laser pulses, and can not be switched again with either a laser pulse or an external magnetic field.
This indicates the high temperature of the sample, and further highlights the importance of proper heat engineering.
\\ \\
To better understand double switching at these ultrashort timescales we turn to modelling, which has been successful in describing AOS in (synthetic) ferrimagnets\cite{atxitia2012llb,Ostler2012GdFeCo,mentink2012theory,wienholdt2013orbital,moreno2017TbCoModel,gridnev2018sdmodel,davies2020pathways}.
Here we use the simplified Microscopic Three Temperature Model (M3TM) as introduced by Beens et al.\cite{beens2019comparing}, which can describe AOS in layered ferrimagnets.
In this model, the system is split in four interacting systems, namely separate spin systems for Co and Gd, mobile spinless electrons, and phonons.
The two spin sublattices carry the magnetization, and are described using a Weiss mean-field approach, whereas the electrons and phonons are described in terms of temperatures.
An incident laser pulse is initially absorbed by the electron system, raising the electron temperature ($T_e$).
Due to electron-phonon scattering, the electron and phonon temperature ($T_p$) will equilibrate.
Ultrafast demagnetization occurs via electron-phonon scattering events, which have a finite probability for an electron to flip its spin.
Exchange of angular momentum between the magnetic sublattices is described using exchange scattering.
The model parameters are taken from Beens et al.\cite{beens2019comparing}.
Similar to previous work\cite{atxitia2018ultrafast}, we include heat diffusion to the substrate by adding a phenomenological term to the phonon temperature, namely
\begin{equation}
\frac{\mathrm{d} T_p}{\mathrm{d} t} \propto \frac{T_{amb} - T_p}{\tau\textsubscript{d}},
\end{equation}
where $T_{amb}$ is the ambient temperature (room temperature), and $\tau\textsubscript{d}$ a characteristic time constant for heat diffusion.
\\ \\
\begin{figure}[t]
    \includegraphics[width=234pt]{{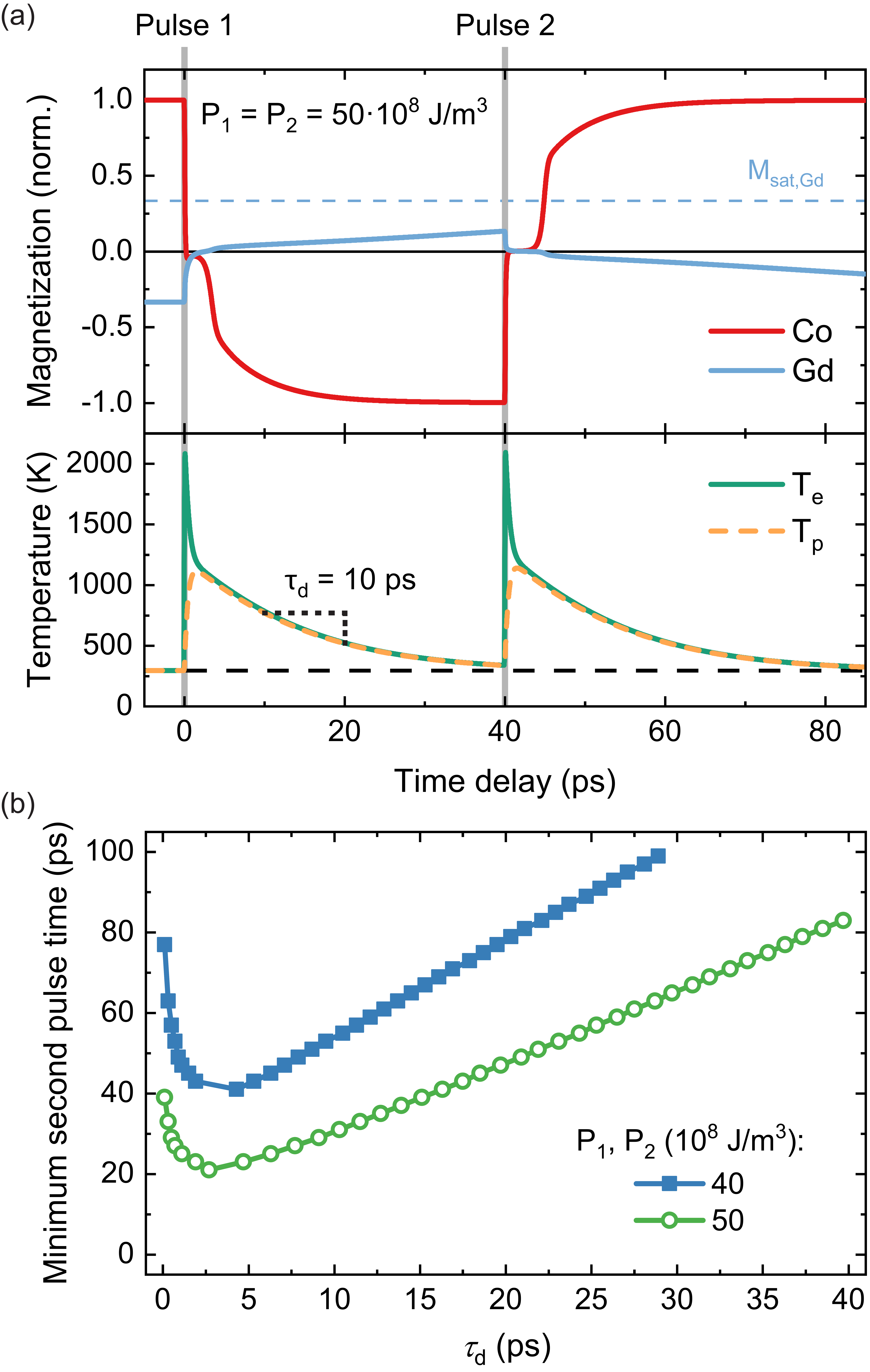}}
    \caption{(a) M3TM simulation results of excitation of a Co/Gd bilayer (3 monolayers each) with two laser pulses (gray lines) separated by 40 ps. The top graph shows the average magnetization response of both Co and Gd over time, with the dashed line indicating the Gd magnetization at saturation. The bottom graph shows the electron and phonon temperature of the combined system, with the ambient temperature $T_{amb}$ (295 K) indicated by the dashed line. (b) Minimum time delay needed between two pulses for consistent double switching as a function of the heat diffusion constant $\tau\textsubscript{d}$, determined via M3TM simulations.}
    \label{fig:fig3}
\end{figure}
Figure \ref{fig:fig3}a shows the simulated magnetization dynamics (top) as well as $T_e$ and $T_p$ (bottom) of a layered Co/Gd system with $\tau\textsubscript{d}$ = 10 ps after excitation with two laser pulses separated by 40 ps.
After the first pulse, the Co and Gd magnetization are switched, followed by remagnetization in the opposite direction.
$T_e$ increases rapidly due to heating by the laser pulse and equilibrates with $T_p$ more slowly, while heat is being removed from the system with the characteristic timescale $\tau\textsubscript{d}$.
As the intrinsic remagnetization rate of Co is faster than $\tau\textsubscript{d}$, the latter is dominant in determining the remagnetization time.
By the time the second pulse arrives, the heat has nearly dissipated from the system, and the Co magnetization is very close to saturation.
For Gd however, the remagnetization rate is an order of magnitude slower\cite{vaterlaus1991spin,koopmans2010paradox,frietsch2020GdTb}, such that Gd has not returned to saturation (dashed line) when the second pulse arrives.
This is in contrast with the expected behavior in alloys, where the Gd has more transition metal neighbors with which it can exchange angular momentum and therefore remagnetize more rapidly.
Nevertheless, this second pulse also leads to the switching of the magnetization of both layers, after which remagnetization towards the original direction proceeds.
Although it might seem counter-intuitive that switching is still possible with a strongly reduced Gd moment, this is in line with previous research on AOS in synthetic ferrimagnets where a change in the relative magnetization of the sublattices was found not to hinder switching\cite{beens2019comparing}.
Conversely, in ferrimagnetic alloys it is known that AOS can only occur if the magnetizations of both sublattices (nearly) cancel each other\cite{xu2017GdFeCoConcentration}.
As such, we conjecture that any possible detriment in remagnetization speed in synthetic ferrimagnets could be compensated by the absence of a need to wait for full magnetization recovery, leading to very similar timescales for repeated AOS.
\\ \\
Finally, we investigate the effect of heat diffusion on double switching in the M3TM.
For different values of $\tau\textsubscript{d}$ we model excitation with two laser pulses.
By varying the time delay between the pulses and evaluating the final magnetization state, we extract a value for the minimum time delay needed for double switching as a function of $\tau\textsubscript{d}$.
Fig. \ref{fig:fig3}b shows these values for two different values of the absorbed laser power density.
For $\tau\textsubscript{d}$ $>$ 5 ps, the minimum waiting time between pulses depends approximately linearly on $\tau\textsubscript{d}$.
In this regime heat diffusion seems to be the dominant factor, however a non-trivial dependence on the power of both pulses is also found.
Comparing the blue squares and green circles, for 40 and \SI{50E8}{\joule\per\metre\cubed} respectively, it is found that the second switch can happen faster if the power of both pulses is higher.
Individual time traces for different pulse powers (not shown here) indicate that the switch occurs faster for higher pulse powers, giving more time for relaxation towards saturation.
For $\tau\textsubscript{d}$ $<$ 5 ps, the minimum time needed for the second pulse starts to increase, which is attributed to a hindrance of the switching mechanism in general by the unrealistically fast dissipation of heat.
Here, the system already cools down enough for the sublattices to start remagnetizing at the timescale at which the switch would normally take place.
From these results it is clear that although efficient heat dissipation is essential for achieving high switching repetition rates, the power also needs to be carefully controlled.
Combining this with the experimental observation that a slight increase in laser power can already be detrimental, this highlights the narrow range of laser powers for which consistent as well as rapid double switching is possible.
Finally, we note that double switching within 10 ps, as observed in \ref{fig:fig1}b, is not reproduced in modelling even for very efficient heat dissipation.
We believe this to be an intrinsic limit of the system modelled here, as this probably represents a minimum time needed for sufficient recovery of the Gd magnetization.
Past work has shown that including intermixing, which is undoubtedly present in the real system, leads to more efficient transfer of angular momentum between the two sublattices\cite{beens2019intermixing}.
Hence, we expect that in an intermixed system both Co and Gd could remagnetize more rapidly than in a bilayer with perfect interfaces, potentially reducing the waiting time needed for the second switch.
It should be noted that switching back within a few picoseconds, as was reported using a different model\cite{atxitia2018ultrafast}, does not seem to be possible using the M3TM for any combination of parameters.
\\ \\
In conclusion, we have investigated the timescales for repeated all-optical switching in synthetic ferrimagnetic Co/Gd bilayers, and have demonstrated a minimum waiting time of 10 ps between two subsequent successful switching events, implying writing speeds of up to 100 GHz.
We have shown that the layered nature of these systems need not be a hindrance to achieve similar writing speeds as in alloys, explained by the notion that the slower remagnetization of Gd is compensated by a less critical dependence of AOS on the Gd moment.
Furthermore, by changing the substrate we have confirmed the importance of engineering heat diffusion away from the magnetic system.
Finally, with modelling efforts using the M3TM we have resolved the role of heat diffusion in ultrafast repeated switching, but we also stress that controlling the laser power is critical to reliable integration in future optically written data storage devices.

\section*{Acknowledgements}
\noindent We gratefully acknowledge M. Beens for assistance on implementation of the M3TM and discussion of simulation results.
This work is part of the Gravitation programme ‘Research Centre for Integrated Nanophotonics’, which is financed by the Netherlands Organisation for Scientific Research (NWO).

\section*{Author declarations}
\subsection*{Conflict of Interest}
The authors have no conflicts to disclose.
\subsection*{Data availability}
The data that support the findings of this study are available
from the corresponding author upon reasonable request.

\clearpage
\bibliography{MainManuscript}

\end{document}